\documentclass[9pt,twocolumn,twoside]{pnas-new}

\templatetype{pnasresearcharticle} 

\usepackage{multirow}
\usepackage[final]{pdfpages}

\begin{document}

\title{Extremal Structures with Embedded Pre-Failure Indicators - Preprint}

\author[a,b,1]{Christoffer Fyllgraf Christensen}
\author[b]{Jonas Engqvist}
\author[a]{Fengwen Wang}
\author[a]{Ole Sigmund}
\author[b]{Mathias Wallin}

\affil[a]{Department of Civil \& Mechanical Engineering, Technical University of Denmark, 2800 Kongens Lyngby, Denmark}
\affil[b]{Department of Construction Sciences, Lund University, SE-22100 Lund, Sweden}

\leadauthor{Christensen}


\authordeclaration{The authors declare no competing interest.}
\correspondingauthor{\textsuperscript{1}To whom correspondence should be addressed. E-mail: chrify@dtu.dk}


\begin{abstract}
    Preemptive identification of potential failure under loading of engineering structures is a critical challenge. Our study presents an innovative approach to built-in pre-failure indicators within multiscale structural designs utilizing the design freedom of topology optimization. The indicators are engineered to visibly signal load conditions approaching the global critical buckling load. By showing non-critical local buckling when activated, the indicators provide early warning without compromising the overall structural integrity of the design. This proactive safety feature enhances design reliability. With multiscale analysis, macroscale stresses are related to microscale buckling stability. This relationship is applied through tailored stress constraints to prevent local buckling in general while deliberately triggering it at predefined locations under specific load conditions. Experimental testing of 3D-printed designs confirms a strong correlation with numerical simulations. This not only demonstrates the feasibility of creating structures that can signal the need for load reduction or maintenance but also significantly narrows the gap between theoretical optimization models and their practical application. This research contributes to the design of safer structures by introducing built-in early-warning failure systems.
\end{abstract}

\dates{This manuscript was compiled on \today}
\doi{\url{}}

\maketitle
\thispagestyle{firststyle}
\ifthenelse{\boolean{shortarticle}}{\ifthenelse{\boolean{singlecolumn}}{\abscontentformatted}{\abscontent}}{}

\firstpage[8]{3}

\noindent
The ability to predict and prevent failure of structures is a crucial safety imperative. It is highly relevant in industries ranging from aerospace through civil engineering to critical infrastructures such as nuclear power plants \cite{McBrearty1956, Bukenya2014, Boller2007, Farrar2007, Giurgiutiu2015, Brownjohn2007}. Generally, two types of failure are distinguished \cite{Yuansheng1988}. The first is local failure due to accidental events resulting in cracks or material loss. This type of failure often comes down to fatigue, corrosion, or collisions. Thus, extensive Structural Health Monitoring (SHM) schemes are enforced to ensure structural integrity throughout the lifetime of a given structure \cite{Balageas2006}. In the aerospace industry, airplanes are examined in specific intervals, and crack growth is monitored to avoid accidents, as even small cracks can result in catastrophic failure \cite{Gallagher1978, Polak2019}. Large structures like bridges are scanned for cracks using drones \cite{Catt2019, Sony2019, Ri2024} to evaluate the current state of the structure. This level of surveillance of the structural state is time-consuming and expensive. Furthermore, it limits knowledge about the state of the structure to the specific time and loading when it is scanned. The second type of failure is due to overloads. This can lead to failure by stressing the material beyond its ultimate tensile strength (UTS) or to a catastrophic collapse by critical buckling of the structure. Strain sensors are commonly used to monitor structural health, but the uncertainty of failure locations often necessitates deploying a large number of sensors. This results in a vast amount of data that must be processed within a limited timeframe \cite{Guemes2024}.

The present work focuses on the second type of failure by embedding pre-failure indicators that warn of overloading. These indicators can potentially also identify the first type of failure, but that is not the scope of the present work. The design approach used to achieve structures with embedded pre-failure indicators in this work is topology optimization \cite{Bendsoe2004}. The method encodes material distribution and geometric variation through computational modeling and optimization to optimize a user-defined objective function subject to one or more constraints.

The work presented here is a new way of handling structural design overload utilizing multiscale material. It incorporates indicators that visibly indicate that loads are approaching the critical state. This is efficiently done through homogenization-based optimization \cite{Wu2021} using knowledge of the relation between macroscale stress states and microscale buckling stability. The objective used in this work is to maximize the critical buckling load of the structure, but other objectives can be used to achieve specific structural properties. The design is optimized while embedding an indicator region showing non-critical local buckling when the load reaches a certain fraction of the critical load. (For full details on the algorithmic approach, see SI, section S1).

The numerically optimized designs are de-homogenized and fabricated using 3D printing with an elastic resin. The specimens are experimentally tested and compared to numerical 3D simulations with good correlation, thus demonstrating the potential for embedded pre-failure indicators in critical structural design.

\section*{Structural Safety using Topology Optimization}
In structural optimization, the primary focus has been on structural efficiency and material minimization. However, work on optimizing fail-safe structures has been initiated during the last ten years. In topology optimization, Jansen et al. \cite{Jansen2014} and the later extension to 3D \cite{Zhou2016} were the first to consider fail-safe structures in the design process. By considering local failure, they produced designs that performed satisfactorily even after one of the structural features failed. An alternative indirect approach was proposed in \cite{Wu2018}, where a maximum feature size was enforced through a local restriction on the amount of material. The outcome was multiscale-like structures, which were insensible to the failure of single features, even without considering failure during optimization. Recently, da Silva et al. \cite{DaSilva2022} incorporated a predetermined failure point for yielding. The approach ensures that pre-failure happens at a location chosen by the engineer while the rest of the structure is strong enough to carry the design load. This approach reduces the need to include a large number of damage scenarios that must be evaluated numerically. Common to the above methods is that they focus on local fracture or yield failure, thus leaving fail-safe designing against overloading without permanent damage a virtually untouched research topic in structural optimization.

\begin{figure*}
    \centering
    \includegraphics[width=17.8cm]{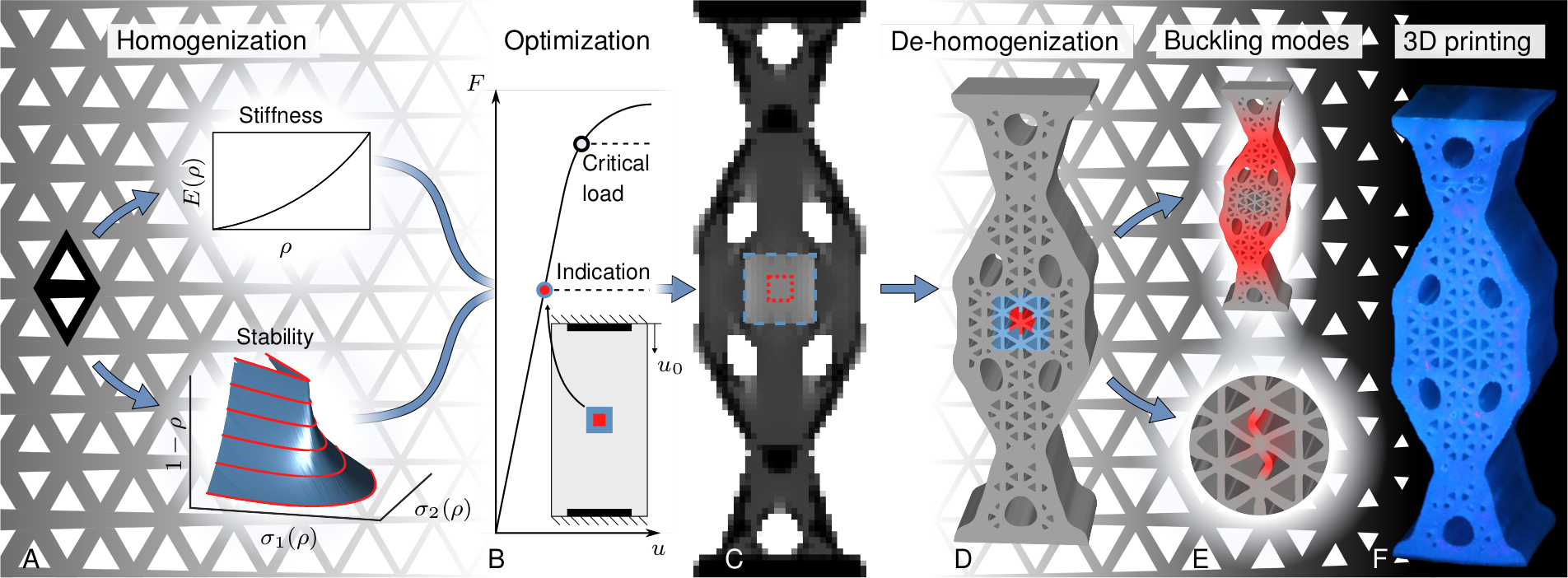}
    \caption{Structural optimization with embedded pre-failure indicator using homogenized isotropic multiscale material. (A) Stiffness and stability data relative to local volume fractions are obtained through homogenization. (B) The optimization is performed on a coarse mesh using the homogenized material properties to enhance computational efficiency. (C) An optimized homogenized design is obtained with the desired buckling response. (D) The optimized design is de-homogenized to extract a high-resolution physical design. (E) The de-homogenized design is post-evaluated numerically. (F) The de-homogenized design is fabricated for experimental validation of the method.}\label{fig:1}
\end{figure*}

With the recent advances in additive manufacturing, multiscale and infill structures have gained increasing interest \cite{Berger2017,Bayat2023}. Work on tailoring microscale properties to achieve specific macroscale material properties is an active research area. Hierarchical structures were designed to maximize the buckling stability of multiscale material in \cite{thomsen2018a}. Lattice structures utilizing self-stress to create localized buckling regions achieving tunable mechanical properties were presented in \cite{Paulose2015}. Similarly, \cite{Injeti2019} examined the use of local internal pre-stress in selected regions for engineering tensile failure load and stiffness in lattice structures. On a single scale level but with multiple materials, \cite{Li2022a} presented a method for programmable load responses. Even though the study focused on singlescale design, the method has potential in multiscale material design. These four examples focus on the geometry and material properties on the microscale level. Utilizing such microstructure material properties in structural design provides multiple advantages. Recent examples are presented in \cite{Christensen2023, Hubner2023, Hubner2023a} where the buckling stability of 2D and 3D multiscale structures is increased using multiscale material.

The embedded pre-failure indicator design procedure used in this work is illustrated in Fig.~\ref{fig:1}. Homogenization determines multiscale material properties such as stiffness and local buckling strength \cite{bensoussan1978a, BENDSOE1988197, Guedes1990a}, Fig.~\ref{fig:1}A. Since the homogenized material properties are independent of design domain and boundary conditions and only depend on the material and geometry of the microstructure \cite{thomsen2018a, Christensen2023}, the homogenization is only performed once. Using the homogenized material properties, a user-defined topology optimization design problem with a chosen indicator region determines the design's shape and the microstructure's local volume fraction, Fig.~\ref{fig:1}B. The indicator is controlled with stress constraints utilizing the relation between macroscale stresses and microscale buckling stability determined using Bloch-Floquet cell analysis \cite{Geymonat1993, Triantafyllidis1985}. It is enforced during optimization using a modified Willam-Warnke failure criterion \cite{Warnke1975, Giraldo-Londono2020}, where the stresses are coupled to the global buckling factor of the structure \cite{Haghpanah2014a, Christensen2023}. This formulation ensures activation at the desired displacement or load, which is specified according to a desired safety factor for the structure. (See SI, section S1, for full details on the optimization procedure). The optimized homogenized design in Fig.~\ref{fig:1}C is de-homogenized to extract a high-resolution physically meaningful design with an embedded pre-failure indicator, Fig.~\ref{fig:1}D. The de-homogenized designs are analyzed in 3D using geometric and material non-linear analysis; see Fig.~\ref{fig:1}E. Finally, the numerical results are compared to experiments performed on 3D printed samples as shown in Fig.~\ref{fig:1}F. (See SI, section S2 for details on numerical evaluations and SI, section S3 for details on experimental tests). 

\section*{Results and Discussions}
Using the method in Fig.~\ref{fig:1}, four structures are designed. The domain, shown in Fig.~\ref{fig:1}B with the indicator region in blue and red, has two black passive solid regions to ensure consistent boundary conditions across homogenized, de-homogenized, and experimental analyses. The design domain is a $6$cm$\times12$cm plane rectangle clamped at the bottom, with a 4cm out-of-plane thickness to prevent out-of-plane buckling. The top edge is vertically displaced by 12mm. The optimizations aim to maximize the Buckling Factor (BF), i.e. increase the displacement at which critical global buckling will occur. Structural compliance is constrained to ensure a minimum stiffness. Finally, the designs are constrained to use less than 30\% of the design domain volume.

All designs are de-homogenized for numerical and experimental tests. This process results in higher material usage than in homogenized structures due to finite periodicity and the addition of boundary shells. Although smaller microstructures can reduce volume discrepancies by more accurately representing the material's volume at boundaries \cite{Christensen2023}, manufacturing constraints limit their practicality. Consequently, the added material in de-homogenized designs increases the BFs. Structural failure in the non-linear analysis and experimental tests is defined as the displacement at which the tangent stiffness is reduced to 70\% of the initial stiffness \cite{Wang2023}.

Two benchmark structures are designed without indicators to show how buckling causes failure in multiscale structures. Next, two designs with embedded pre-failure indicators are optimized to demonstrate the potential and advantage of early warning systems in the design process.

\subsection*{Uniaxial Compression Benchmark Designs}
%
\begin{figure*}
    \centering
    \includegraphics[width=17.8cm]{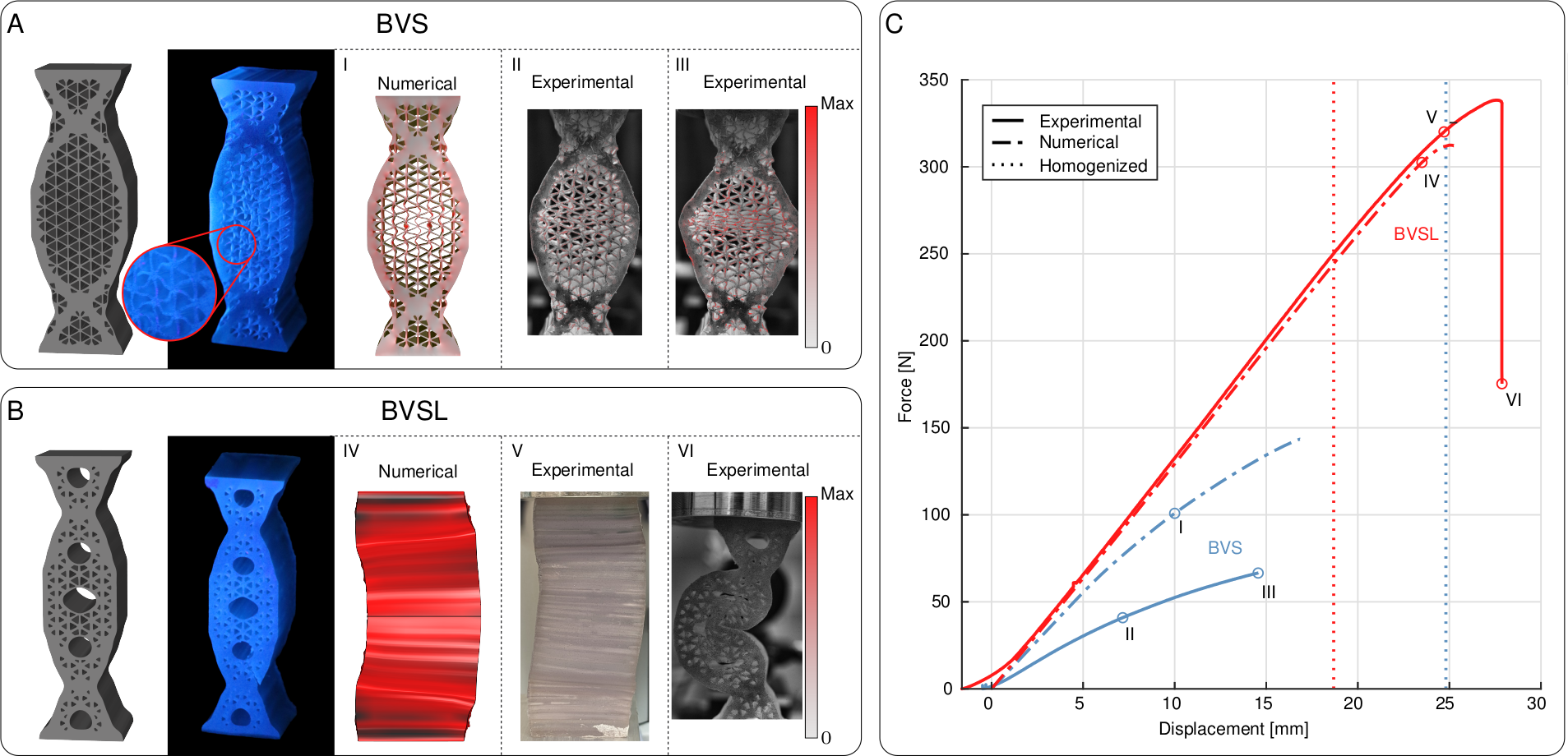}
    \caption{Optimized designs without the pre-failure indicator. (A) Numerical and printed BVS designs optimized for maximizing global buckling (The picture of the printed structure is taken with UV light to enhance contrast). The design is sensitive to local buckling, even from residual stresses resulting from the printing process, as seen in the unloaded printed sample. I--II shows the numerical and experimental displacements for the BVS design at 70\% stiffness. III shows the local buckling at the maximum experimental displacement of the BVS design. The color bar represents the 2D in-plane von Mises stress (arbitrary units). (B) BVSL structure optimized for maximizing global buckling while preventing local buckling of the microstructure. The picture is taken after the experiment, hence the fracture in the bottom part of the structure. IV--VI shows the displacements of numerical and experimental analysis for the BVSL design. The color bar in IV represents the 2D out-of-plane von Mises stress (arbitrary units). (C) Load/displacement response for numerical and experimental tests with failure points indicated by circles at I, II, IV and V. Vertical dotted lines indicate the critical displacements estimated by the homogenized analysis.}\label{fig:2}
\end{figure*}

The first design, denoted BVS (Buckling, Volume, Stiffness, with the first letter, B, indicating the objective of the optimization problem and the remaining indicating active constraints, thus not including local buckling stability), is seen in Fig.~\ref{fig:2}A. 

The de-homogenized structure has a significantly lower critical BF than the homogenized one, as shown in Tab.~\ref{tab:1}. Fig.~\ref{fig:2}C illustrates the stiffness differences and early softening observed in both numerical analyses and experimental tests. This lower BF and early softening occur because the de-homogenization breaks the separation of scales assumption, introducing a risk of local buckling. The very slim features of the BVS design reduce its load-carrying capacity, as seen in Fig.~\ref{fig:2}AI--III. This sensitivity to local buckling is also evident in the 3D-printed sample, which shows local buckling throughout the microstructure due to residual stresses from the printing process, as shown in Fig.~\ref{fig:2}A.

\begin{table}[t!]
    \centering
    \caption{Homogenized, de-homogenized and experimental performance data for the four optimized designs}\label{tab:1}
    \begin{tabular}{lrrrr}
       & \textbf{BVS} & \textbf{BVSL} & \textbf{BVSLI} & \textbf{BVSLID} \\
    \midrule
    \multicolumn{5}{c}{Homogenized Analysis} \\
    \midrule
    Volume fraction & 0.3 & 0.3 & 0.3 & 0.3 \\
    Global BF & 2.07 & 1.56 & 1.44 & 1.30 \\
    \midrule
    \multicolumn{5}{c}{De-homogenized 3D Non-Linear Buckling Analysis} \\
    \midrule
    Volume fraction & 0.322 & 0.322 & 0.328 & 0.332 \\
    Critical BF & 0.846 & 1.93 & 1.85 & 1.81 \\
    Indicator BF & - & - & 0.816 & 0.737 \\
    Relative indication & - & - & 44.1\% & 40.7\% \\
    \midrule
    \multicolumn{5}{c}{Experimental Analysis} \\
    \midrule
    Critical BF & 0.597 & 2.06 & 2.13 & 1.82 \\
    Indicator BF & - & - & 0.844 & 0.677 \\
    Relative indication & - & - & 39.6\% & 36.6\% \\
    \bottomrule
    \end{tabular}
    
\end{table}

The second design, denoted BVSL with L for Local stability, is shown in Fig.~\ref{fig:2}B. It is designed using the same objective and constraints as the BVS design but with the addition of a constraint to prevent critical local buckling of the microstructure; see SI, section S1A. The performance of the BVSL design is also provided in Tab.~\ref{tab:1}. Comparing the de-homogenized and printed designs show immediate improvements over the BVS design, as local buckling is not present before the structure is loaded. The load/displacement response shown in Fig.~\ref{fig:2}C is nearly linear up to and beyond the critical displacement estimated by the homogenized analysis. Furthermore, the numerically predicted stiffness is very close to the experimentally observed stiffness when compensating for manufacturing defects; see SI, section S3A. Evaluating the buckling modes in Fig.~\ref{fig:2}BIV--VI confirms that the structure fails due to global buckling and not local buckling. However, the initial buckling mode is primarily in the out-of-plane direction despite the thick extrusion, as seen in Fig.~\ref{fig:2}BIV and V. When the printed structure is loaded beyond the initial failure, it snaps to the in-plane mode in Fig.~\ref{fig:2}BVI resulting in a rapid drop in the load/displacement response revealing a dramatic and critical failure with little to no warning in advance.

The failure modes of the BVS and BVSL designs and their sensitivity to small overloads highlight the necessity to consider buckling on all scales and motivate the use of embedded pre-failure indicators.

\subsection*{Embedded Pre-Failure Indicator Designs}
Two versions of the embedded pre-failure indicator method are investigated. The first design, denoted BVSLI with I for Indicator, uses the full domain. The union of the blue and red squares is the region where the local buckling constraint is relaxed to allow for non-critical local buckling to be triggered at 35\% of the global BF. An additional constraint enforcing local buckling at 40\% of the global BF is included in the red square. This constraint is equivalent to a safety factor of 2.5 from indication to catastrophic failure.

The second design, denoted BVSLID with D for Damaged, optimizes for the physical state problem in the full domain equivalent to the BVSLI design, while also including a damaged version of the domain; see SI, section S1B for more details. Removing the damaged red and blue squares from the state analysis ensures that the global structure remains stable even if the load-carrying capability of the indicator region is lost. This approach introduces a conservative way of accounting for the softening effects in the indicator after activation.
~
\begin{figure*}
    \centering
    \includegraphics[width=17.8cm]{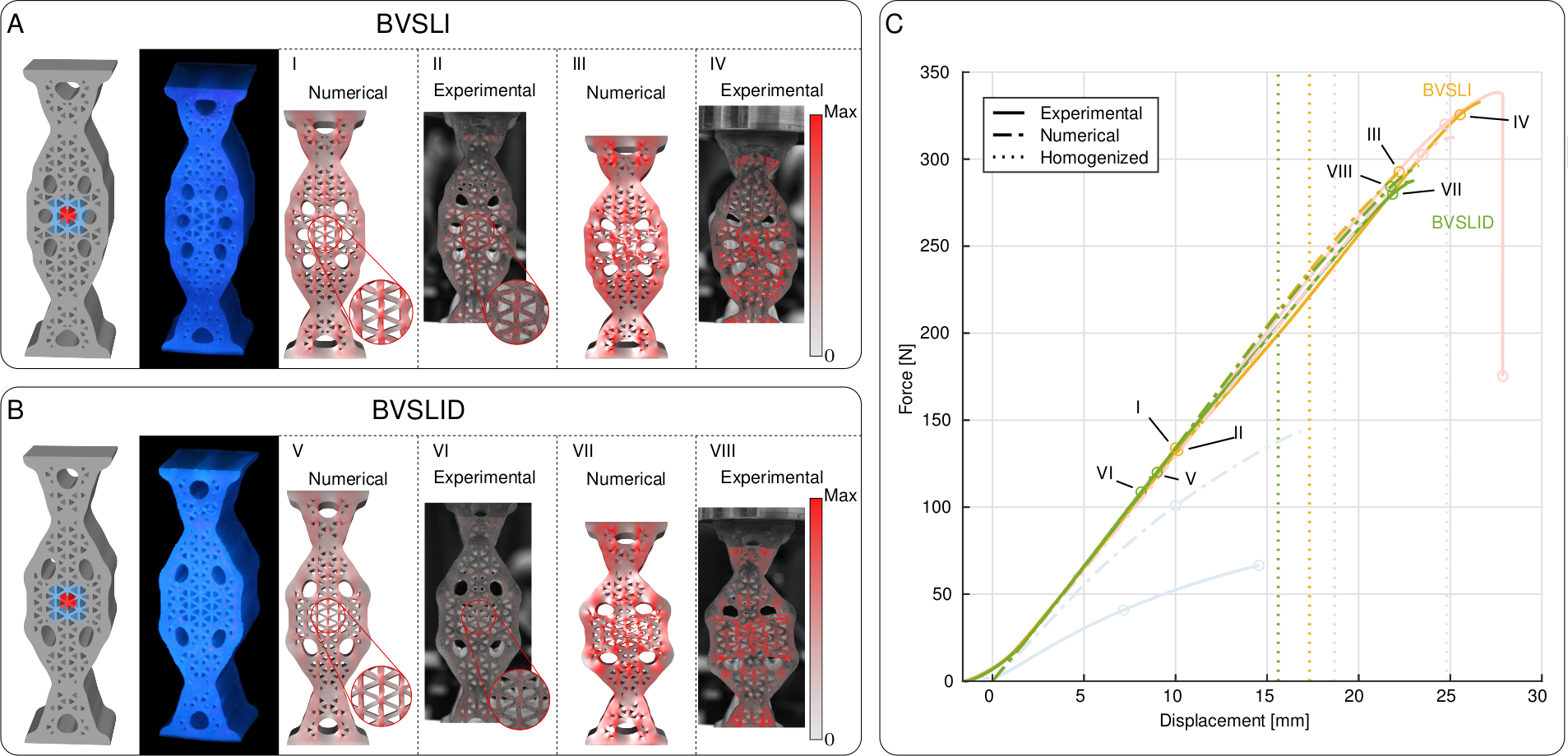}
    \caption{The two optimized designs with embedded pre-failure indication. (A) Numerical and printed BVSLI designs optimized for maximizing global buckling while embedding a pre-failure indicator. I--IV presents the related buckling modes with the color bar showing the 2D in-plane von Mises stress (arbitrary units). (B) BVSLID structure optimized for maximizing global buckling while accounting for local buckling of the indicator. (C) Load/displacement response for the numerical and experimental tests with an indication of indicator activation and failure points by circles.}\label{fig:3}
\end{figure*}

The BVSLI design is presented in Fig.~\ref{fig:3}A. Comparing it to the BVSL design in Fig.~\ref{fig:2}B, it is seen that they have similar dense microstructures. At the same time, the indicator region in the BVSLI design is slightly less dense making it more sensitive to buckling as desired. As for the BVS and BVSL designs, the volume of the final de-homogenized design is slightly higher than the homogenized designs, see Tab.~\ref{tab:1}. The table also shows the BF evaluated at the critical displacement and the displacement at which the indicator is activated. 

The non-linear analysis shows an excellent correlation between the local and global buckling modes compared to the experimental results; see Fig.~\ref{fig:3}AI--IV. The numerical load/displacement response in Fig.~\ref{fig:3}C also matches the experimental response well. The activation of the indicator in the numerical analysis is determined, equivalent to the global failure criterion, as the displacement at which the tangent von Mises stress is 30\% higher than the initial, see SI, section S2B. This approach is not possible in the experimental analysis due to the gradual initiation of contact between the test sample and the load cell. Instead, a polynomial is fitted to the von Mises stress obtained using Digital Image Correlation (DIC). The initiation of local buckling is determined at the displacement where the gradient of the von Mises stress changes the fastest.

The BF for activation of the indicator is very similar for the experimental and numerical analysis as seen in Tab.~\ref{tab:1}. However, the critical failure of the structure, estimated from the numerical analysis, occurs significantly later than the anticipated BF derived from the homogenized analysis. The numerical analysis of the structure indicates global failure at a BF of 1.85. This results in an indicator activation at 44\% of the global BF, thus reducing the safety factor to 2.27. It also means that the critical BF estimated in the numerical analysis is 17\% lower than the experimental results. The deviation can be attributed to stiffening effects due to self-contact within the indicator region. The difference between activation of the indicator and global failure in the experiment is 40\%, putting it within the target of $35\%$--$40\%$ as required in the optimization. However, as indicated by the numerical analysis, if self-contact is not contributing to stability, critical failure will occur earlier, such that activation of the indicator is outside the target window. This effectively means that the performance of the design depends on self-contact, which is not included in the optimization.

To handle the uncertainties associated with the large deformations in the indicator region, we consider the worst-case BVSLID approach. This approach results in the design in Fig.~\ref{fig:3}B, with the performance data in Tab.~\ref{tab:1}. 
As a result of the de-homogenization, the BFs of the numerical analysis are higher than those estimated by the homogenized analysis. The numerical analysis also shows a good correlation with the experimental test. The pre-failure indicator and critical global modes are presented in Fig.~\ref{fig:3}BV--VIII. As for the BVSLI design, the correlation between the numerical and experimental deformations is very good. The load/displacement response in Fig.~\ref{fig:3}C shows excellent matches in terms of stiffness and the critical global failure point as the numerically predicted BF is 1.81 which should be compared to 1.82 in the experiment. This is indeed satisfactory and shows that the design does not depend on self-contact to perform at the expected level. The pre-failure indicator is activated at a BF of 0.74 in the numerical simulation resulting in activation at 41\% of critical failure. The pre-failure indicator is experimentally activated at a BF of 0.68, i.e. 37\% of the critical failure which is inside the target window.

In summary, the BVSLID design is buckling resistant beyond the estimated values in the homogenized 2D optimization, and it does not rely on self-contact. It has an embedded pre-failure indicator experimentally validated to be activated at approximately 37\% of the critical global buckling point and shows an almost linear stiffness response up until the point of critical global buckling. Based on this, the BVSLID design successfully fulfills the initial requirements stated for the optimization problem, while showing a good correlation between numerical and experimental results.

\subsection*{Concluding Remarks}
Multiple structures were optimized with the objective of maximizing the critical buckling factor. Two designs were used as benchmarks, highlighting the level of previous state-of-the-art multiscale optimization methods for maximizing buckling stability. Two additional designs were optimized to show the potential of embedded pre-failure indicators. All designs were numerically and experimentally post-evaluated with good correlation to confirm the validity of the proposed method. The validation confirmed that it is possible to utilize topology optimization of multiscale structures with tailored local buckling stress constraints to develop and design extremal structures with embedded pre-failure indicators as an early warning of catastrophic failure. The method has potential extensions that are interesting to investigate in future studies. The present study focuses on the scenario of failure due to overloading. However, it can potentially also warn against failure due to cracks or material loss and this would be an obvious study direction. Additional possible extensions include multiple materials to allow stiff structures with flexible indicators. The present work used one indicator with one indicator target, but the method can easily be extended to any number of pre-failure indicators.  Each indicator has the possibility to warn against individual overload scenarios or, potentially, other failure scenarios.

\matmethods{Details of the topology optimization with embedded pre-failure indicators are provided in SI, section S1. Details on the numerical post-evaluations are available in SI, section S2. The experimental setup is described in SI, section S3.}


\showmatmethods{} 

\acknow{The authors acknowledge funding from Villum Fonden through the Villum Investigator Project “InnoTop”.}

\showacknow{} 

\newpage
\bibsplit[21]

\bibliography{MyPnas}

\newpage


\section*{Supplementary material}

**Supporting Information Text**

**1. Topology optimization with embedded pre-failure indicators using multiscale material**

The section presents the topology optimization of multiscale material and embedded pre-failure indicator framework. The method is built on top of the topology optimization of multiscale structures framework in Christensen et al. (1). For this reason, only a brief description of underlying material interpolations and multiscale buckling stability modeling will be presented here. The section is divided into three subsections: the definition of multiscale modeling and related interpolations; the definition of the pre-failure indicator; and the definition of the optimization problems.

**A. Definition of multiscale modeling and interpolations.** The pre-failure indicator method utilizes the relation between microscale buckling stability and macroscale stresses following the principles described in (1–3). This is achieved through Bloch-Floquet wave analysis, which effectively retains the analysis of the repetitive unit cell of the microstructure while capturing all wavelengths of instability under imposed periodicity boundary conditions. A multifield method is used to allow holes and to control the minimum feature size in the design (4). The design is defined by the density field $x_j$, which controls the relative density of the microstructure and the indicator field $s_j$ controlling where the material is allowed. These two fields are combined to formulate the physical design field $\rho_j$

$$\rho_j = \tilde{x}_j \bar{\tilde{s}}_j, \tag{1}$$

where $\tilde{x}_j$ is a smoothed version of $x_j$ and $\bar{\tilde{s}}_j$ is a smoothed and projected version of $s_j$. The full definition of the multifield method and the filtering and projection settings can be found in (1).

In this work, an isotropic triangular microstructure is used. Therefore, the Hashin-Shtrikman (HS) upper bound for isotropic materials is applied to interpolate the bulk $\kappa$ and shear $\mu$ moduli of the multiscale structure.

$$\mu(\rho_j) = \rho_j \mu_0 + (1-\rho_j)\mu_{\min} - \frac{\rho_j(1-\rho_j)(\mu_0 - \mu_{\min})^2}{(1-\rho_j)\mu_0 + \rho_j\mu_{\min} + \frac{\kappa_0\mu_0}{\kappa_0 + 2\mu_0}},$$

$$\kappa(\rho_j) = \rho_j \kappa_0 + (1-\rho_j)\kappa_{\min} - \frac{\rho_j(1-\rho_j)(\kappa_0 - \kappa_{\min})^2}{(1-\rho_j)\kappa_0 + \rho_j\kappa_{\min} + \mu_0}, \tag{2}$$

where $\rho_j$ is the density of the $j$–th element. The upper ($\mu_0, \kappa_0$) and lower ($\mu_{min}, \kappa_{min}$) bound of the bulk and shear moduli are determined by the Elastic 50A resin from Formlabs used for 3D printing. The stiffness of this material was estimated in (5) as $E_0 = 2.736$MPa, $E_{min} = E_0 \times 10^{-6}$ and a constant Poissons ratio $\nu_0 = \nu_{min} = 0.3$ is used.

The upper limits for the microstructure buckling strength are determined using Bloch-Floquet wave theory (2, 6) at eight different relative densities equally distributed in the interval $[0.1, 0.8]$. Based on these limits, a Willam-Warnke failure surface combined with the interpolation function in Eq. 3 from (7) is fitted to formulate a smooth density-dependent buckling failure surface in terms of the macroscale stresses.

$$\bar{\sigma}_k(\tilde{x}_j) = E_0 \left( b_{0,k}\tilde{x}_j^{n_0} + b_{1,k}\tilde{x}_j^{n_0+1} \right), \quad k \in \{c, t, b\} \tag{3}$$

where $\tilde{x}_j$ is element microstructure density and $b_{0,k}$, $b_{1,k}$ and $n_0$ are the fitted coefficients whose value are stated in Tab. S1. The coefficients are fitted to uniaxial compression $c$, uniaxial tension $t$ and biaxial compression $b$. To alleviate numerical instability for high densities, where local buckling is highly unlikely, Eq. 3 is relaxed above densities of 0.8 using a smooth heaviside function $\mathcal{H}$ (8), which yields

$$\sigma_k(\tilde{x}_j) = \bar{\sigma}_k(\tilde{x}_j) + \psi \mathcal{H}(\tilde{x}_j, \eta, \bar{\beta}). \quad k \in \{c, t, b\}. \tag{4}$$

In Eq. 4, $\psi$ is a relaxation parameter, which in this work is 10 times the stress limit for a solid unit cell. The sharpness parameter is $\bar{\beta} = 50$ and the threshold $\eta$ is defined at the highest density of the known data points i.e. $\eta = 0.8$.

The Willam-Warnke failure surface is defined by

$$\sigma_{eq}(\rho_j, \tilde{x}_j) = \alpha(\rho_j, \tilde{x}_j)\sqrt{3J_2(\rho_j)} + G(I_1(\rho_j), \tilde{x}_j), \tag{5}$$

where $J_2$ is the second invariant of the deviatoric stress tensor **s** and $I_1$ is the first invariant of the Cauchy stress tensor $\boldsymbol{\sigma}$. The term $G(I_1(\rho_j), \tilde{x}_j)$ is determined from the first invariant of the Cauchy stress, see (1). Finally, the shape of the failure surface is defined by $\alpha(\rho_j, \tilde{x}_j)$ as

$$\alpha(\rho_j, \tilde{x}_j) = \frac{A(\tilde{x}_j)\cos^2(\hat{\theta}(\rho_j)) + B(\tilde{x}_j)}{C(\tilde{x}_j)\cos(\hat{\theta}(\rho_j)) + \sqrt{D(\tilde{x}_j)\cos^2(\hat{\theta}(\rho_j)) + E(\tilde{x}_j)}}, \tag{6}$$

with $A(\tilde{x}_j)$, $B(\tilde{x}_j)$, $C(\tilde{x}_j)$, $D(\tilde{x}_j)$ and $E(\tilde{x}_j)$ determined from the stress limits and $\hat{\theta}(\rho_j^m)$ being the modified Lode angle. Refer to (1) for details regarding the multiscale buckling stress constraint formulation.

**B. Definition and modeling of pre-failure indicators.** During optimization the macroscale problem is solved using linearized buckling analysis (9)

$$\mathbf{K}(\boldsymbol{\rho})\mathbf{u}_0 = \mathbf{f}_0, \qquad [7]$$

$$[\mathbf{G}_\sigma(\boldsymbol{\rho}, \mathbf{u}_0) + \gamma_i \mathbf{K}(\boldsymbol{\rho})]\boldsymbol{\varphi}_i = \mathbf{0}, \qquad [8]$$

where Eq. 7 is the linear elastic state problem with reference load $\mathbf{f}_0$ and reference displacement $\mathbf{u}_0$. The problem is solved using the density-dependent stiffness matrix $\mathbf{K}(\boldsymbol{\rho})$. Eq. 8 is the linearized buckling eigenvalues problem with the eigenpairs $(\lambda_i, \boldsymbol{\varphi}_i)$, $i \in \mathcal{B}_0$ representing the eigenvalues and buckling mode vectors. The buckling factor (BF) is obtained by $\lambda_i = 1/\gamma_i$. The stress stiffness matrix $\mathbf{G}_\sigma(\boldsymbol{\rho}, \mathbf{u}_0)$ is computed using the density $\boldsymbol{\rho}$ and stresses $\boldsymbol{\sigma}_0$ corresponding to the reference displacement $\mathbf{u}_0$. The critical stresses $\boldsymbol{\sigma}^{crit}$ are obtained from the linearized buckling assumption.

$$\boldsymbol{\sigma}^{crit} = \lambda_{crit}\boldsymbol{\sigma}^0 \qquad [9]$$

where $\boldsymbol{\sigma}^0$ are the macroscale stresses at the reference load $\mathbf{f}_0$ and $\lambda_{crit}$ is the global buckling factor. This method ensures buckling stability on all scales without performing high-resolution FEA in every design iteration. The current study utilizes the stress scaling in Eq. 9 to embed pre-failure indicators at a user-defined location in a given design domain. An illustration of a domain and indicator region is presented in Fig. S1. Following the procedure from (10), two density representations are used. The first is the nominal design in Fig. S1A. The second is the damaged design in Fig. S1B. The damaged design aims to introduce a conservative way of accounting for the softening effect of local buckling in the indicator during activation.

The nominal design is split into three subsets to incorporate the pre-failure indicator. The first set is the external region, the second is the indicator relaxation region, and the final is the indicator region, illustrated in Fig. S1C–E, respectively. These sets are used when enforcing stress constraints to obtain the desired microscale buckling behavior. The size of the indicator relaxation region should be large enough to allow an upper bound on the stresses in the external region that is lower than the lower bound on the stresses in the indicator region. In this work, the side length of the indicator relaxation region is $l^{rel} = l^{ind} + 2r_s$, where $l^{ind}$ is the side length of the indicator region and $r_s$ is the filter radius used on the topological indicator field $s$, see (1).

All element indices in the design domain belong to the set $\mathbb{N}_k$. The indices in the external, relaxed, and indicator regions are included in $\mathbb{N}_k^{ext}$, $\mathbb{N}_k^{rel}$, and $\mathbb{N}_k^{ind}$, respectively. The following set rules apply: $\mathbb{N}_k^{ext} \cup \mathbb{N}_k^{rel} = \mathbb{N}_k$ and $\mathbb{N}_k^{ind} \subset \mathbb{N}_k^{rel}$. The damaged design is obtained by assigning the void indicator field in the elements in $\mathbb{N}_k^{rel}$ to $\mathbf{s} = \mathbf{0}$ before filtering and projection is performed.

The external region is subject to the local buckling constraints stated in (1) with a stress modification such that Eq. 9 is scaled with a factor $\alpha_U^{ext} \leq 1$.

$$\boldsymbol{\sigma} = \alpha_U^{ext}\lambda_{crit}\boldsymbol{\sigma}^0. \qquad [10]$$

The local buckling constraint is relaxed in the indicator relaxation region by lowering stresses with a factor $\alpha_U^{rel} < \alpha_U^{ext} \leq 1$ to allow local buckling earlier in this region. In a case where only the critical BF $\lambda_{crit}$ is considered, the relaxed stresses are calculated as

$$\boldsymbol{\sigma}^{rel} = \alpha_U^{rel}\lambda_{crit}\boldsymbol{\sigma}^0, \qquad [11]$$

where $\boldsymbol{\sigma}^{rel}$ are the relaxed stresses and $\alpha_U^{rel}$ is the relaxation factor.

The indicator set has an additional lower bound on the stresses to ensure buckling when the load exceeds a predefined portion of the global critical load. This behavior is controlled by the factor $\alpha_L$ such that $\alpha_U^{rel} < \alpha_L < \alpha_U^{ext} \leq 1$. It is used in the same way as $\alpha_U^{rel}$ in Eq. 11 to calculate $\boldsymbol{\sigma}^{ind}$.

In Eq. 9–11 $\lambda_{crit}$ is the minimum BF. Since the minimum of all $\lambda_i$ $i \in \mathcal{B}_0$ is non-differentiable, we approximate the min-function by the Kreisselmeier-Steinhauser (KS) function (11). In practice, only a subset of eigenpairs $\mathcal{B} \subset \mathcal{B}_0$ are included in the KS function. The KS function provides a smooth upper-bound approximation of the minimum eigenvalue $\lambda_1$.

**C. Multiscale optimization with embedded pre-failure indicators.** Four design optimizations are performed to illustrate and motivate the use of the proposed embedded pre-failure indicators. The robust formulation (8) is used to formulate an *eroded*, *blueprint* and *dilated* design, $m \in \{e, b, d\}$. The overall optimization problem is formulated as

$$
\begin{aligned}
\min_{\mathbf{x},\mathbf{s}} \quad & : \quad \max_m \left( g_{BF}(\boldsymbol{\rho}^{m,n}) \right), & m \in \{e,b,d\},\ n = \{1,2\}, \\
\text{s.t.} \quad & : \quad g_s(\boldsymbol{\rho}^{e,1}) = 1 - \frac{\mathbf{u}(\boldsymbol{\rho}^{e,1})^T \mathbf{K}\mathbf{u}(\boldsymbol{\rho}^{e,1})}{S_e^*} \leq 0, \\
& : \quad g_V(\boldsymbol{\rho}^{d,1}) = \frac{\sum_j v_j \rho_j^{d,1}}{V_d^* V_\Omega} - 1 \leq 0, \\
& : \quad g_l^k(\boldsymbol{\rho}^{m,n}, \tilde{\boldsymbol{x}}^n, \gamma_i(\boldsymbol{\rho}^{m,n}), \alpha_U^k) \leq 0, & \forall k \in \mathbb{N}_k,\ m \in \{e,b,d\},\ n = \{1,2\},\ \gamma_i \in \mathcal{B} \\
& : \quad -g_l^k(\boldsymbol{\rho}^{m,1}, \tilde{\boldsymbol{x}}^1, \gamma_i(\boldsymbol{\rho}^{m,1}), \alpha_L) \leq 0, & \forall k \in \mathbb{N}_k^{ind},\ m \in \{e,b,d\},\ \gamma_i \in \mathcal{B} \\
& : \quad \rho_j^{m,n} = \tilde{x}_j \bar{\tilde{s}}_j^{m,n}, & \forall j \in \mathbb{N}_j,\ m \in \{e,b,d\},\ n = \{1,2\} \\
& : \quad x_{min} \leq x_j \leq 1, & \forall j \in \mathbb{N}_j \\
& : \quad 0 \leq s_j \leq 1. & \forall j \in \mathbb{N}_j
\end{aligned}
\qquad [12]
$$

where $\boldsymbol{\rho}^{m,n}$ are the physical densities associated with the robust fields $m$ and the $n^{\text{th}}$ density realization; $n = 1$ is the nominal design and $n = 2$ is the damaged design. The physical density field is constructed by the filtered density field $\tilde{\mathbf{x}}$ and the void indicator field $\bar{\tilde{\mathbf{s}}}^{m,n}$ following (4). The objective $g_{BF}(\boldsymbol{\rho}^{m,n})$ is to maximize the BFs using the KS function $J^{KS}$

$$g_{BF}(\boldsymbol{\rho}^{m,n}) = \frac{J^{KS}\left(\gamma_i(\boldsymbol{\rho}^{m,n})\right)}{J_0^{KS}}, \quad m \in \{e, b, d\}, \ n = \{1, 2\}, \ \gamma_i \in \mathcal{B}. \quad [13]$$

The objective is normalized with the initial value of the KS function $J_0^{KS}$. A total of 12 buckling modes are included in the KS function with an aggregation parameter of 100. A stiffness constraint $g_s$ is enforced on the eroded design with the lower limit $S_e^*$. The volume constraint $g_V$ is enforced on the dilated design with a maximum allowed volume fraction $V_d^*$. The dilated volume fraction is updated continuously to ensure that the target volume $V_b^*$ is enforced on the blueprint design following the strategy from (8) to eliminate numerical artifacts and stabilize the convergence of the optimization problem. The design domain presented in the main article is rectangular with a horizontal side length being half of the height. To avoid boundary effects the domain in the optimization is square with both sides equal to the height. Thus $V_b^*$ is adjusted with the actual area of the optimization design domain. The local buckling stress constraint $g_l^k$ is enforced on the entire design domain with the scale factors $\alpha_U^k$ defined according to the sets $\mathbb{N}_k^{ext}$ and $\mathbb{N}_k^{rel}$. The indicator constraint $g_i^k$ is enforced in the indicator set $\mathbb{N}_k^{ind}$ with the lower bound stress factor $\alpha_L$. All constraint values are presented in Tab. S2 with $V_b^*$ defined for the domain in the main article.

## 2. Numerical post-evaluation of designs

The designs obtained by the optimization problem in Eq. 12 are post-processed and evaluated. First, de-homogenization is used to obtain physical interpretations of the homogenized designs. Second, the designs are numerically evaluated in Comsol 6.1.

**A. De-homogenization of optimized designs.** The optimized designs are de-homogenized following the method from (12) with the closing shell suggested in (1). The cell size is chosen based on experience with the Formlabs Form 3+ 3D printer. This means the lattice size between two parallel bars in the microstructure is 4.8mm or approximately four elements in the homogenized design.

The orientation of the microstructure is chosen based on the macroscale principal stresses. The purpose is to ensure the desired buckling response in the indicator region. This necessity is visible in Fig. S2B, where the relation between macroscale stresses and microscale buckling stability is illustrated. The Willam-Warnke failure surface, used to ensure the local buckling stability, is fitted to the worst case of all cell orientations. This means that the choice of orientation influences how well and conservatively the local buckling stability is prevented. In a design case where preventing buckling is the only objective, this conservative relation is straightforward and requires no special consideration during de-homogenization. However, in the present work, where non-critical local buckling is desired in the indicator region, the accuracy of the fit is crucial.

For this reason, the principal stresses are examined in the designs. In Fig. S2A, the BVSLID design is presented with the principal stresses indicated. Based on the stresses in the indicator region, the orientation selected for de-homogenization is $\theta = 30°$ such that one bar in the microstructure is oriented along the vertical direction.

**B. Numerical buckling analysis.** The de-homogenized designs are numerically evaluated in Comsol 6.1. The evaluation consists of two parts. First, linearized buckling analysis is used to calculate BFs for all designs. Second, a non-linear analysis is conducted with perturbations based on the linearized buckling analysis. All numerical post-evaluations are performed in 3D to get the best correlation with the experimental tests.

The numerical analysis uses quadratic Lagrange brick elements to mesh the geometry. To prevent out-of-plane buckling, the structures are extruded 40mm. The material used for 3D printing is the Elastic 50A from Formlabs. This material's parameters are determined in (5), i.e., it is modeled using an incompressible Neo-Hookean material model with Young's modulus $E_0 = 2.736$MPa.

For the linearized buckling analysis, a linear elastic model with Poisson's ratio $\nu_0 = 0.495$ is used. The reference displacement used in the linearized buckling analysis is $u_0 = 12$mm, corresponding to the homogenized optimization. Selected local and global buckling modes from the linearized buckling analysis are presented in Fig. S3. These modes are used to introduce imperfections in the otherwise symmetric designs when performing the non-linear buckling analysis.

The non-linear analysis uses the incompressible material model from (5) with a mixed formulation for better numerical conditioning. Furthermore, the HuHu regularisation method from (13, 14), is used to reduce higher-order deformation of the elements. The regularization is used with $15^{\text{th}}$ order Gaussian quadrature to get quadrature points as close as possible to the edges (The high integration order could be replaced by exact Gauss-Lobatto integration (15), but this is not available in Comsol 6.1). The regularization and integration order help stabilize the system by hindering element inversion, thus improving convergence. The regularization consists of adding a term to the strain energy density

$$\Psi = \bar{\Psi} + k_r e^{-5|F|} \mathbb{H}u \vdots \mathbb{H}u, \quad [14]$$

where $\bar{\Psi}$ is the strain energy density of the Neo-Hookean material model. The amount of regularization is scaled through the scalar $k_r$, which in this work is $k_r = 10^{-6} E_0$. The determinant of the deformation gradient is $|F|$. The regularization itself is penalized through the Hessian of the displacement field $\mathbb{H}u$.

The critical displacement is determined at the point where the tangent stiffness is reduced to 70% of the initial stiffness (5). Equivalently, the displacement at which the indicator is activated is determined where the maximum tangent von Mises stress in the indicator region is 30% higher than the initial maximum tangent von Mises stress as illustrated in Fig. S4. For comparison with the experiments, where only planar displacements are measured, the 2D von Mises stress is used.

## 3. Experimental validation

The de-homogenized designs are fabricated and tested experimentally to validate the proposed method for embedded pre-failure indicators. This section describes the fabrication process and the introduction of related deviations from the ideal digital designs. Furthermore, the experimental setup is described.

**A. Fabrication of structures.** The structures are fabricated using Formlabs Form 3+ SLA 3D printer. The material used is the Elastic 50A resin. After printing, the structures are rinsed by hand with isopropyl alcohol and compressed air. Second, they are rinsed in an ultrasound shaker with isopropyl alcohol for 10 minutes before being rinsed in the Form Wash for 10 minutes. Finally, the structures are cured in the Form Cure UV light chamber for 30 minutes at 60° with 405nm UV light.

The fabrication process introduces some defects to the otherwise perfectly symmetric designs. The two most significant defects are out-of-plane height deviations and leftover resin in long small extruded holes. The significance of these defects is moderate but introduces some deviations that are quantified here.

First, the out-of-plane height deviations are shown for the BVSLI design in Fig. S5A. The differences in height for all the designs are presented in Tab. S3, where the maximum height deviation in the print direction is 1.5mm. This difference leads to more straining on one side than the other. Thereby, a perturbation of the out-of-plane buckling mode is introduced and the risk of activating it as the first global mode is increased. Fig. S5B shows out-of-plane buckling of the BVSL structure shortly before the onset of the in-plane buckling mode. The out-of-plane deviation also results in an initial gradual establishment of contact visible in the experimental load/displacement response in Fig. S6A. The experimental responses are adjusted to account for this gradual contact establishment. This is done by fitting a $7^{th}$ order polynomial to the experimental data and using the largest gradient to extrapolate linearly to estimate the equivalent contact point. The correction values are reported in Tab. S3 under *Gradient correction*. The correction values show a good correlation with the out-of-plane height difference. Correcting the load/displacement response provides a better comparison to numerical results.

Second, the residual resin in the printed structure increases the stiffness. The printed structures' weights are compared to the expected weights based on the volume and a material density of 1088kg/m$^3$ to quantify this effect. The data is presented in Tab. S3 and shows an increase in weight of around 6–8%. The increased amount of material will increase the stiffness of the structures. Assuming a linear relation between weight and stiffness, this is accounted for by correcting the gradient in the numerical load/displacement curves with the respective weight deviation. Fig. S6 compares the experimental and numerical load/displacement curves before and after adjusting the stiffness according to the increased weights. The adjusted curves in Fig. S6B show a better fit of the numerical data to the experimental than the unadjusted curves in Fig. S6A.

**B. Experimental validation.** All fabricated structures were tested up to the critical buckling point. This was done by compressing each sample using a Shimadzu EZ-S test frame with a custom-made control system. The axial load and the displacement were measured using a 500N capacity load cell (model SM-500N, Interface, Inc., US) and a 100mm capacity linear displacement sensor (model WA100mm, HBM, Germany), respectively. The measured force and displacement were recorded continuously during the tests that were performed using a constant displacement rate of 5 mm/min, measured on the moving grip of the load frame.

To measure the local deformation of the specimen, 2D-surface DIC (Digital Image Correlation) was performed using two 29-megapixel digital cameras (model Prosillica GT6600, Allied Vision Technologies, Germany) that acquired images of the front and back side of the specimen during the compression test. A fine speckle pattern of black, water-based paint was applied to each of the two surfaces to facilitate the DIC. The image correlation was performed using the commercial software Vic-2D 6 (Correlated Solutions, Inc., USA) with a correlation window size of $29 \times 29$ pixels and a step size of 7 pixels.

The global failure point is determined based on the load/displacement response as the point where the tangent stiffness is reduced to 70% of the initial stiffness. The point where the pre-failure indicator is activated was determined by fitting a $7^{th}$ order polynomial to the maximum von Mises strain in the indicator region. Given the out-of-plane height difference, the initial maximum tangent von Mises strain can not be determined. For this reason, the activation criterion used in the numerical analysis is not possible to use for the experimental data. Instead, the activation of the pre-failure indicator is determined at the point where the tangent von Mises stress changes the fastest (3), i.e. where the $3^{rd}$ derivative of the polynomial is equal to zero, as visualized in Fig. S7.

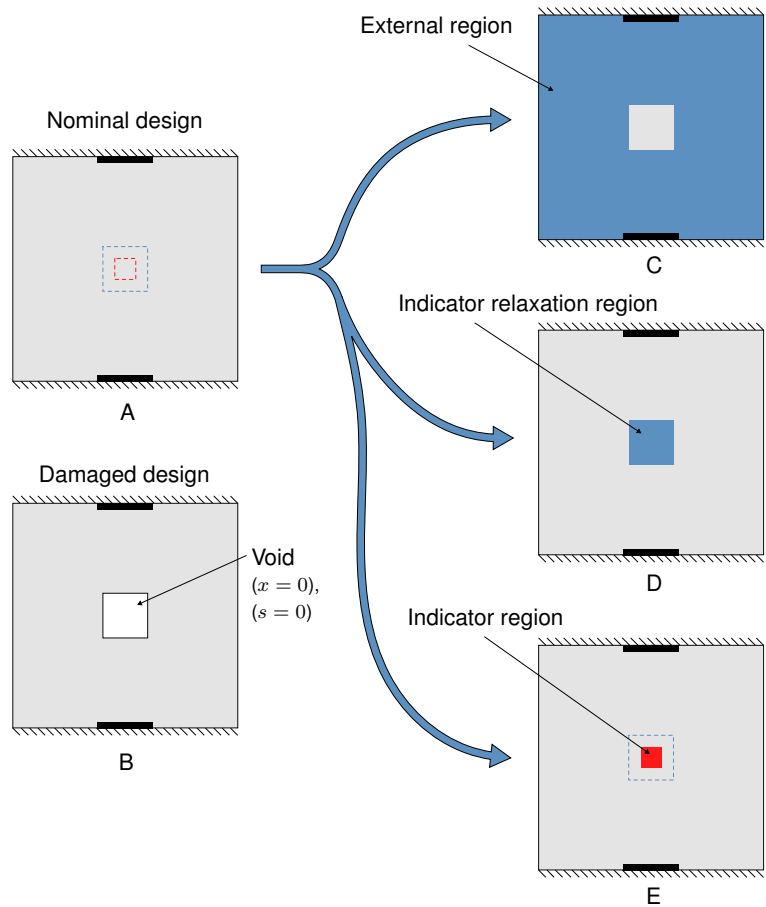

**Fig. S1.** Design domain with an indication of: (A) The nominal design without damage; (B) The damaged design (with $x = 0$ and $s = 0$ in the indicator region). The nominal design is split into three parts: (C) the external region; (D) The indicator relaxation region; (E) The indicator region.

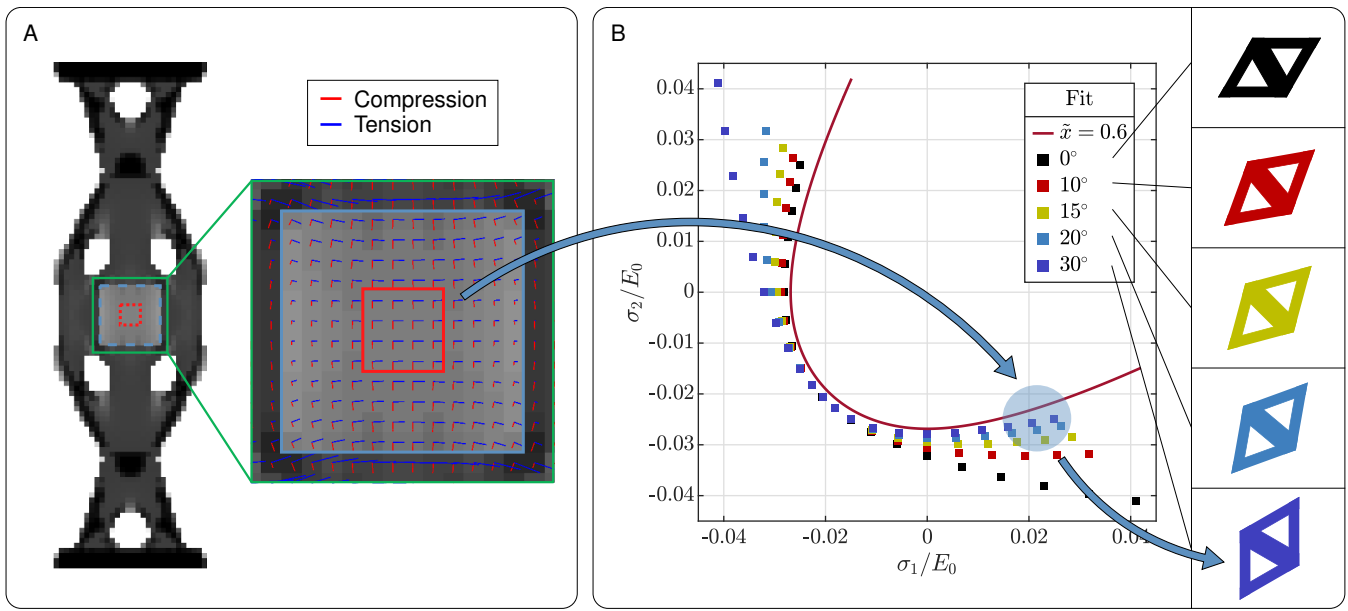

**Fig. S2.** Illustration of the influence of the cell orientation in de-homogenization. (A) The BVSLID design with the local principal stresses plotted on top to determine the optimal cell orientation. (B) Buckling stress limits for different cell orientations and the Willam-Warnke failure surface fitted to the worst case of these orientations.

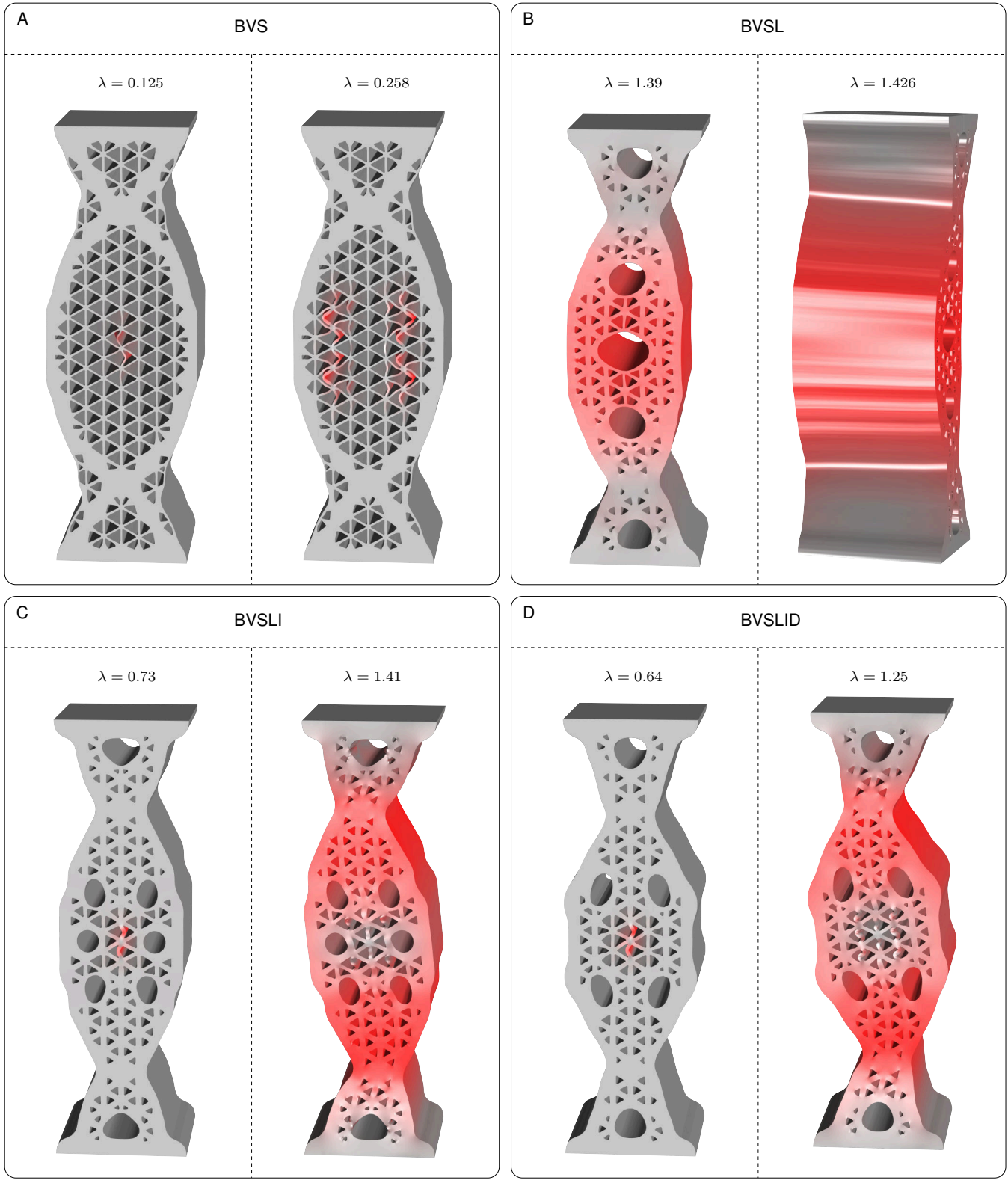

**Fig. S3.** Linearized buckling modes of de-homogenized designs. (A) The first two buckling modes of the BVS design, are both local modes. (B) The first two buckling modes of the BVSL design, where both modes are global and the second mode is in the out-of-plane direction. (C) Indicator and global failure mode of the BVCLI design. (D) Indicator and global failure mode of the BVCLID design.

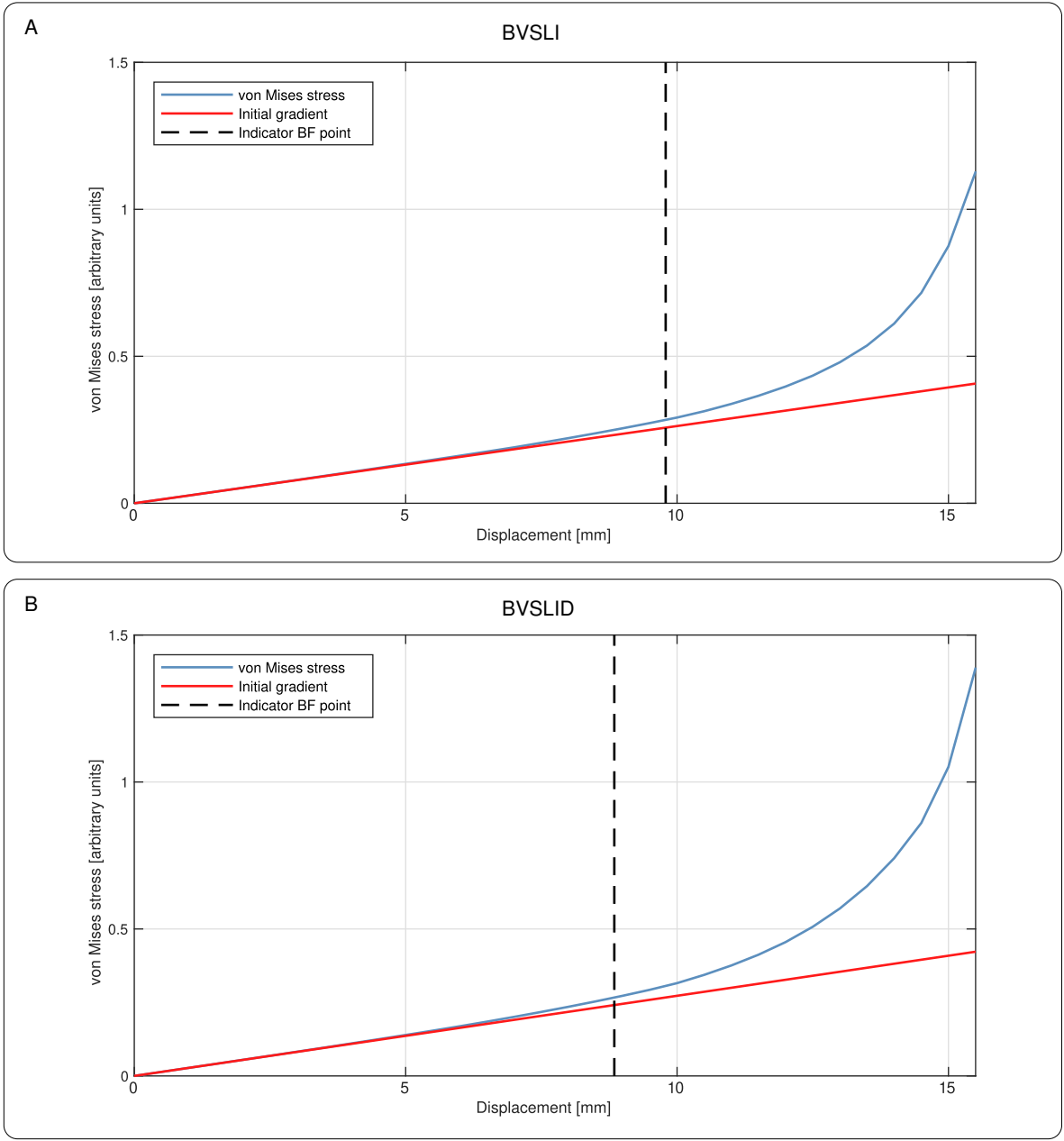

**Fig. S4.** Determination of indicator activation in the numerical analysis using von Mises strain in the indicator. (A) von Mises strain and indicator BF point for the BVSLI design. (B) von Mises strain and indicator BF point for the BVSLID design.

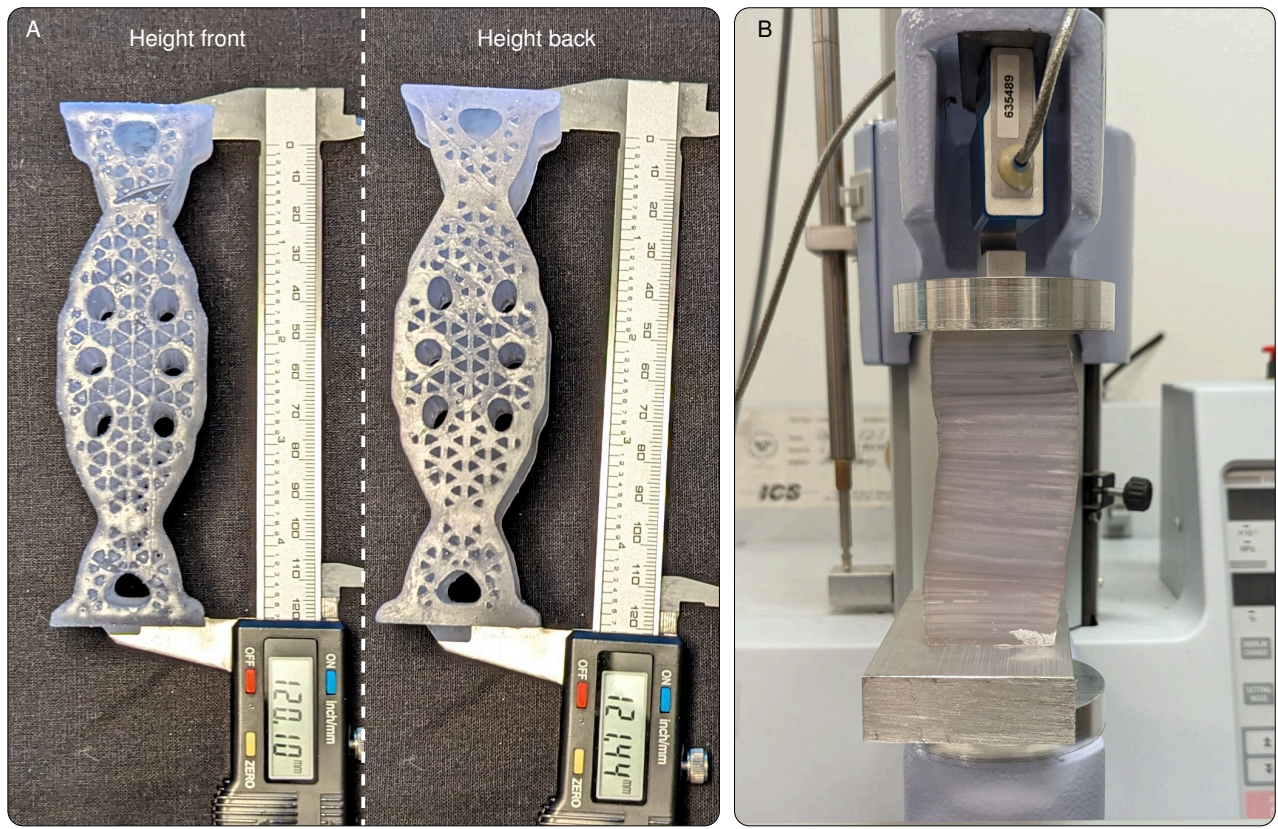

**Fig. S5.** Examples of fabrication defects. (A) showing variation in height in the out-of-plane direction. (B) Out-of-plane buckling of the BVSL structure shortly before the onset of the in-plane buckling mode.

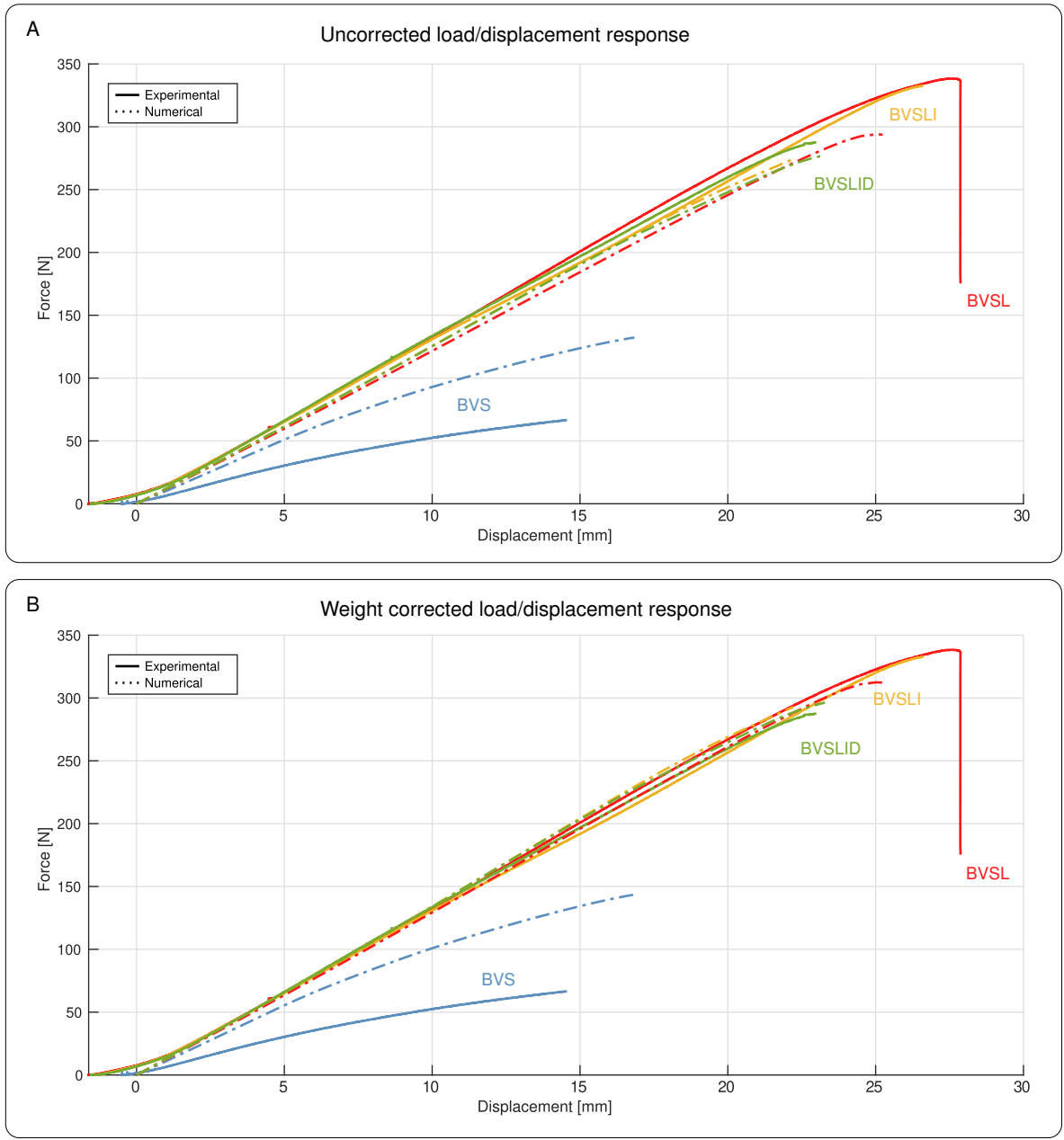

**Fig. S6.** Correction of stiffness according to weight deviations. (A) Uncorrected load/displacement response. (B) Corrected load/displacement response.

Christoffer Fyllgraf Christensen, Jonas Engqvist, Fengwen Wang, Ole Sigmund and Mathias Wallin

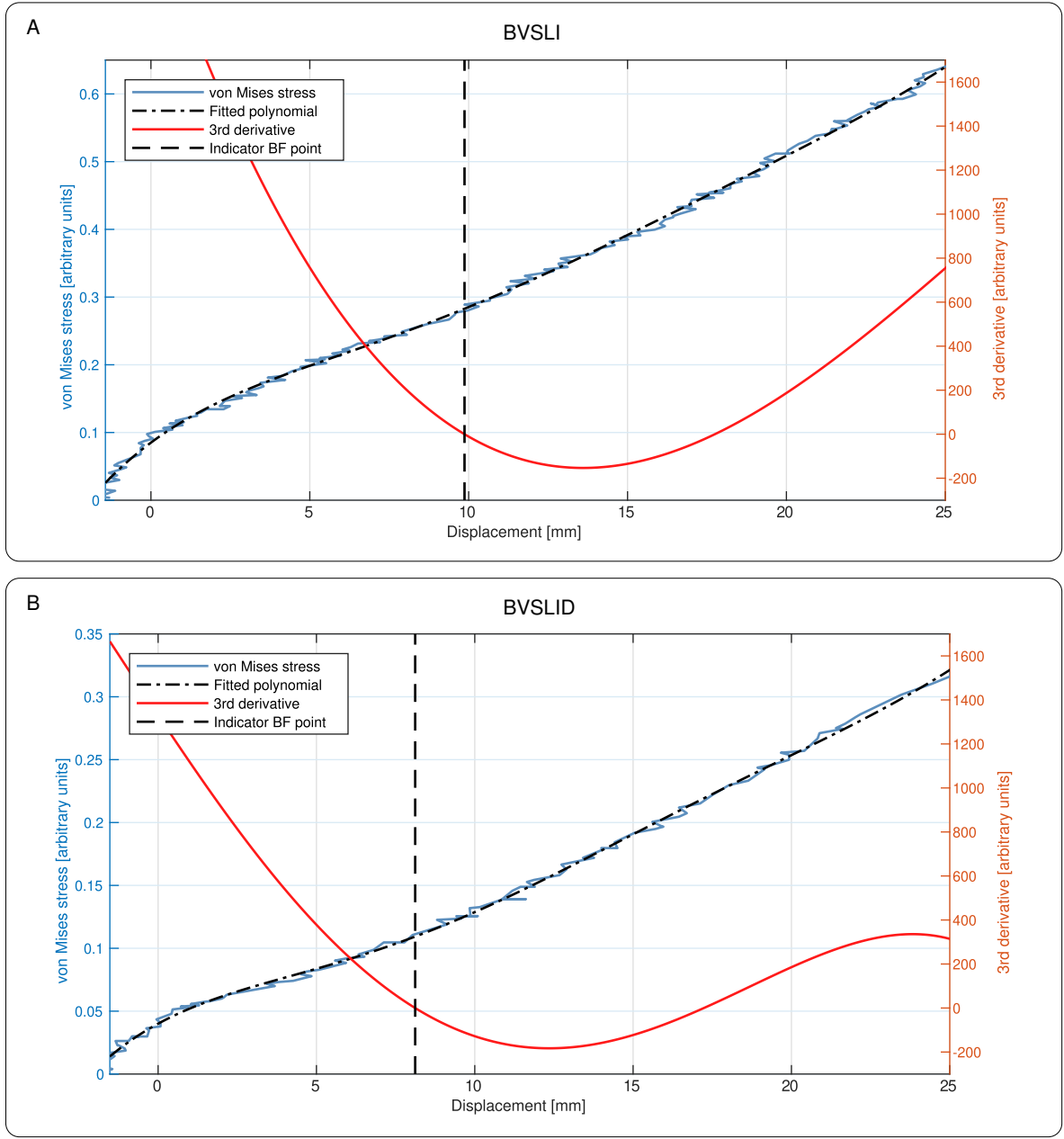

**Fig. S7.** Determination of indicator activation in the experimental analysis using von Mises strain in the indicator. (A) von Mises strain and indicator BF point for the BVSLI design. (B) von Mises strain and indicator BF point for the BVSLID design.

**Table S1. Values of the fitted coefficients used to describe the Willam-Warnke failure surface.**

|         | $c$     | $t$        | $b$     |
|---------|---------|------------|---------|
| $b_{0,k}$ | 0.05882 | 0.3327     | 0.05079 |
| $b_{1,k}$ | 0.1092  | 3.301e$-$13 | 0.06596 |

      **Christoffer Fyllgraf Christensen, Jonas Engqvist, Fengwen Wang, Ole Sigmund and Mathias Wallin**

**Table S2. Optimization parameters**

| $V_b^*$ | $S_e^*$ | $\alpha_U^{ext}$ | $\alpha_U^{rel}$ | $\alpha_L$ |
|---|---|---|---|---|
| 0.15 | $2.758 \times 10^{-1} Nm$ | 1 | 0.35 | 0.4 |

**Table S3. Overview of manufacturing dimension and weight deviations**

| Design | Height front | Height back | Height difference | Gradient correction | Expected weight | Measured weight | Error |
|---|---|---|---|---|---|---|---|
| BVS | 120.24mm | 120.75mm | 0.51mm | 0.49mm | 100.897g | 109.508g | 8.53% |
| BVSL | 120.09mm | 121.57mm | 1.48mm | 1.62mm | 100.897g | 107.251g | 6.3% |
| BVSLI | 120.10mm | 121.44mm | 1.34mm | 1.43mm | 102.777g | 109.640g | 6.68% |
| BVSLID | 119.77mm | 121.19mm | 1.42mm | 1.52mm | 104.030g | 110.982g | 6.68% |

## References

1. CF Christensen, F Wang, O Sigmund, Topology optimization of multiscale structures considering local and global buckling response. *Comput. Methods Appl. Mech. Eng.* **408**, 115969 (2023).
2. CR Thomsen, F Wang, O Sigmund, Buckling strength topology optimization of 2D periodic materials based on linearized bifurcation analysis. *Comput. Methods Appl. Mech. Eng.* **339**, 115–136 (2018).
3. GL Bluhm, O Sigmund, F Wang, K Poulios, Nonlinear compressive stability of hyperelastic 2D lattices at finite volume fractions. *J. Mech. Phys. Solids* **137**, 103851 (2020).
4. R Giele, JP Groen, N Aage, CS Andreasen, O Sigmund, On approaches for avoiding low-stiffness regions in variable thickness sheet and homogenization-based topology optimization. *Struct. Multidiscip. Optim.* **64**, 39–52 (2021).
5. F Wang, M Brøns, O Sigmund, Non-Hierarchical Architected Materials with Extreme Stiffness and Strength. *Adv. Funct. Mater.* **33** (2023).
6. N Triantafyllidis, M Schraad, Onset of failure in aluminum honeycombs under general in-plane loading. *J. Mech. Phys. Solids* **46**, 1089–1124 (1998).
7. MN Andersen, F Wang, O Sigmund, On the competition for ultimately stiff and strong architected materials. *Mater. Des.* **198**, 109356 (2021).
8. F Wang, BS Lazarov, O Sigmund, On projection methods, convergence and robust formulations in topology optimization. *Struct. Multidiscip. Optim.* **43**, 767–784 (2011).
9. RD Cook, DS Malkus, ME Plesha, RJ Witt, *Concepts and Applications of Finite Element Analysis*. (Wiley), Fourth edition, pp. 179–198 (2002).
10. GA da Silva, AT Beck, O Sigmund, Structural topology optimization with predetermined breaking points. *Comput. Methods Appl. Mech. Eng.* **400**, 115610 (2022).
11. G Kreisselmeier, R Steinhauser, Systematic Control Design by Optimizing a Vector Performance Index. *IFAC Proc. Vol.* **12**, 113–117 (1979).
12. JP Groen, FC Stutz, N Aage, JA Bærentzen, O Sigmund, De-homogenization of optimal multi-scale 3D topologies. *Comput. Methods Appl. Mech. Eng.* **364**, 112979 (2020).
13. GL Bluhm, Internal contact modeling for finite strain topology optimization. *Comput. Mech.* **67**, 1099–1114 (2021).
14. AH Frederiksen, O Sigmund, K Poulios, Topology optimization of self-contacting structures. *Comput. Mech.* **-** (2023).
15. O Faltus, M Horák, M Doškář, O Rokoš, Third Medium Finite Element Contact Formulation for Pneumatically Actuated Systems. *ArXiv* **2405.01185** (2024).



\end{document}